# Brownian Motion of Boomerang Colloidal Particles

Ayan Chakrabarty[1], Andrew Konya[1], Feng Wang[1], Jonathan V. Selinger[1]*, Kai Sun[2], Qi-Huo Wei[1]*
[1]Liquid Crystal Institute, Kent State University, Kent, OH 44242
[2] Department of Materials Science and Engineering, University of Michigan, Ann Arbor, MI 48109, USA

We investigate the Brownian motion of boomerang colloidal particles confined between two glass plates. Our experimental observations show that the mean displacements are biased towards the center of hydrodynamic stress (CoH), and that the mean-square displacements exhibit a crossover from short time faster to long time slower diffusion with the short-time diffusion coefficients dependent on the points used for tracking. A model based on Langevin theory elucidates that these behaviors are ascribed to a superposition of two diffusive modes: the ellipsoidal motion of the CoH and the rotational motion of the tracking point with respect to the CoH.

Brownian motion as a general phenomenon of the diffusion processes has inspired extensive research [1-12] due to both its interesting physics and practical applications such as in microrheology [13-16], self-propelled microswimmers [17] and particle and molecular separation [18-20]. Inspired by the diverse geometric shapes of biological macromolecules, Brenner and others have extended the hydrodynamic theory of Brownian motion to particles with irregular shapes [21-26]. A set of hydrodynamic centers are introduced, which include the center of hydrodynamic stress (CoH) where the coupling diffusion matrix becomes zero, the center of reaction where the coupling resistance matrix becomes symmetric and the center of diffusion where the coupling diffusion matrix becomes symmetric [22, 24, 27]. For screw-like or skewed particles, the translational and rotational motions are intrinsically coupled, therefore, the CoH does not exist and the centers of diffusion and reaction differ from each other. By contrast, for non-skewed particles, there always exists a unique point at which these three hydrodynamic centers coincide.

Thus far, experimental studies of Brownian motion have been focused primarily on spherical particles; it was only recently that the Brownian motion of low-symmetry particles was explored in experiments [5, 28-34]. Particle shapes are critical to various applications such as self-propelled microswimmers and particle/molecular separations [17, 35]. By engineering particle shapes, microswimmers may be tailored to perform circular, spinning-top or other types of motion [35-37]. Understanding the hydrodynamics of chiral particles may lead to new avenues towards separation of particle or molecular enantiomers [38].

In this letter we study the Brownian motion of boomerang-shaped colloidal particles under quasi two-dimensional confinements. The boomerang particles with $C_{2v}$ mirror symmetry represent an attractive system for studying the Brownian motion of low symmetry particles because their CoM and CoH do not coincide and both lie outside the body. Especially, the location of the CoH is unknown before the motion of any tracking point (TP) is analyzed. Boomerang particles are also an interesting model system for active microswimmers [37], the electro-optical properties of DNA molecules [39-42] and the liquid crystal ordering [43].

Our experimental and theoretical studies show that the diffusion of the boomerangs is rather different from that of spheres and ellipsoids. (1) The mean displacements (MD) for fixed initial angle are biased towards the CoH, and the mean square displacements (MSDs) exhibit a crossover from short to long time diffusion with different diffusion coefficients. (2) The boomerangs confined in quasi-two dimensions are non-skewed and possess a CoH where translation and rotation are decoupled in the body frame. (3) Our model based on Langevin theory shows that the non-zero MDs result from the Brownian orbital motion of the TP with respect to the CoH. (4) Two methods for calculating body frame displacements which give indistinguishable results for ellipsoids, yield drastically different results for boomerang particles.

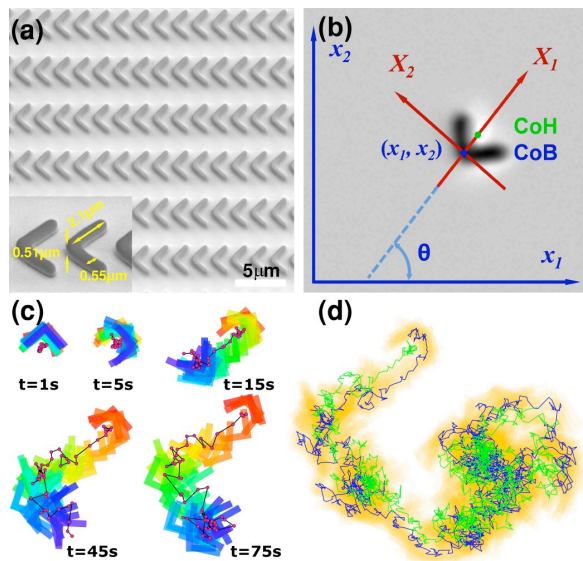

Fig. 1. (a) SEM image of the boomerang particles fabricated on silicon wafer. (b) Optical microscopic image and schematics of the coordinate systems. ($x_1$-$x_2$): the lab frame and ($X_1$-$X_2$): the body frame. (c) Representative trajectories of 5 different total lengths of time, where the red spots represent the positions of the CoH, and the boomerang is colored-coded in time. (d) An exemplary 300s trajectory for the CoH (green) and the CoB (blue).

The boomerang colloidal particles made of photo-curable polymer (SU8) were fabricated by using photolithography [44] and have a 2.1 μm arm length, 0.51 μm thickness, 0.55 μm arm width and 90° apex angle [Fig. 1(a)]. The aqueous suspension of the particles, stabilized by adding sodium dodecyl sulfate, (SDS, 1mM) was filled in a cell of ~ 2 μm thickness. Videos of isolated moving boomerangs were taken using a CCD camera at time step $\tau$ = 0.05 s. Limited by the computer memory, each video contains 3000 frames and a total of 167 videos were taken for the same particle.

We developed an image processing algorithm to track the position and orientation of the boomerang particles. The cross-point between the central axes of the two arms represents the center of the body (CoB) and is a convenient point for motion tracking. The angle bisector gives the particle orientation $\theta$ [Fig. 1(b)]. The precision of our optical microscope and tracking algorithm is determined to be ±13nm and ±0.004 rad for position and orientation respectively. Trajectories obtained from all videos were merged into a single trajectory of ~ $5\times10^5$ frames. To merge two trajectories together, the particle coordinates in the second video are shifted and rotated such that the position and orientation of the first frame in the second video matches those of the last frame in the first video. Figs. 1(c-d) show representative trajectories of the CoB and CoH at different time scales where the coupling between translational and rotational motions can be easily observed (the method to locate the CoH will be discussed later).

In Fig. 2(a), the angle averaged MSDs of the CoB along $x_1$ and $x_2$ are identical, implying that the Brownian motion is isotropic on average. In contrast to that the MSDs for ellipsoids grow linearly with time, the MSDs for the boomerangs exhibit linearity with time only in short and long times with a nonlinear crossover region around $t$ = 10 s. Best linear fittings give the short- and long-time diffusion coefficients respectively as $\overline{D}^{ST}$ = 0.082 μm$^2$/s, $\overline{D}^{LT}$ = 0.057 μm$^2$/s. In Fig. 2(b), The rotational Brownian motion is linear for all times $\langle[\Delta\theta(t)]^2\rangle = 2D_\theta t$, with the diffusion coefficient $D_\theta$ = 0.044 rad$^2$/s.

To discern anisotropic features in the Brownian motion, we measured the MSDs with the initial angle fixed at $\theta_0$=0. Due to its anisotropic shape, the MSDs at short times exhibit different diffusion coefficients along $x_1$ and $x_2$. At long times when the directional memory is washed out, the MSDs grow again linearly with $t$ with identical slope for both $x_1$ and $x_2$ [Fig. 2(c)].

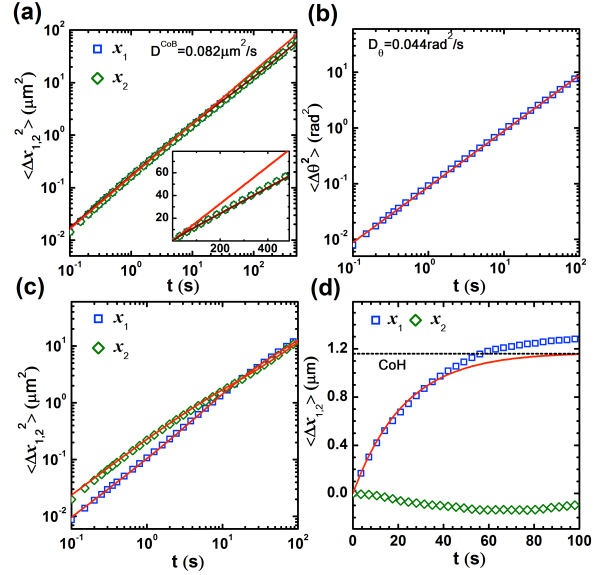

**Fig. 2.** (a) MSDs of the CoB in the lab frame vs. $t$. Red line: the best linear fit for $t$ < 10s; dark brown line: theoretical fit using Eq. 2. Inset: linear plot of MSDs vs. $t$. (b) MSDs of $\theta$ vs. $t$ with the best linear fitting (red line). (c) MSDs for the CoB in the lab frame with $\theta_0$ =0. Red lines: theory curves with Eq. 1b. (d) MDs in the lab frame with $\theta_0$ =0. Red line: theory curve of Eq. 1a using $D_\theta$ and $D_{2\theta}$ obtained from Fig. 2b and 3e.

To note, the MSDs along $x_1$ is larger than that along $x_2$ at long times. As a comparison, MSDs for ellipsoids are identical along $x_1$ and $x_2$ at long times [5].

Although the MDs for Brownian motion are typically zero, we find that it is not the case for the boomerangs. The MDs averaged over different initial angle $\theta_0$ are indeed zero (data not shown here). However, with initial angle fixed at $\theta_0$=0, the MDs along $x_1$ are non-zero [Fig. 2(d)] and saturate at long times. Such non-zero MDs along the symmetric line are in sharp contrast with the zero MDs observed for spheres and ellipsoids.

To understand these observations, we assume that the boomerangs confined in 2D possess a CoH (as will later be proved experimentally). The position of a TP on the symmetric line can be expressed as: $\boldsymbol{x}(t) = \boldsymbol{x}^{CoH}(t) - r\cos\theta(t)\hat{\boldsymbol{x}}_1 - r\sin\theta(t)\hat{\boldsymbol{x}}_2$, where $\boldsymbol{r}$ is the vector linking the CoH to the TP. When the CoB is used as the TP, we denote the CoB-CoH separation as $d_0$. From the definition of the CoH, the descriptions of the Brownian motion of the CoH requires only one rotation diffusion coefficient $D_\theta$ and two translation diffusion coefficients $D_{22}^{CoH}$ and $D_{11}^{CoH}$, and therefore the Langevin equations for the CoH are actually the same as those for an ellipsoid. As shown in the supplementary material [45], the MDs and MSDs of the TP for fixed initial angle $\theta_0$ can be written as:

$$\langle\Delta x_i(t)\rangle_{\theta_0} = ra_i\tau_1(t) \qquad (1a)$$

$$\left\langle [\Delta x_i(t)]^2 \right\rangle_{\theta_0} = 2\overline{D}^{CoH} t + \cos 2\theta_0 (r^2/2 + \qquad (1b)$$

$$\Delta D/4D_\theta) b_i \tau_4(t) + 2r^2 a_i^2 \tau_1(t)$$

where $\tau_n = 1 - exp(-nD_\theta t)$, $a_1 = \cos\theta_0$, $a_2 = \sin\theta_0$, $b_1 = -1$, $b_2 = 1$, $\overline{D}^{CoH} = (D_{11}^{CoH} + D_{22}^{CoH})/2$, $\Delta D = D_{22}^{CoH} - D_{11}^{CoH}$. Eq. 1a indicates that $\left\langle \Delta x_1(t) \right\rangle_{\theta_0}$ for the CoB saturates at $r = d_0$ in the long time, or the Brownian motion is biased towards the CoH. These theoretical expressions agree well with the experimental results [Fig. (2c-d)].

Averaging Eq. 1a and 1b over different initial angle $\theta_0$ leads to that the angle averaged MDs are zero and the angle-averaged MSDs are expressed as [45]:

$$\left\langle [\Delta x_{1,2}(t)]^2 \right\rangle = 2\overline{D}^{CoH} t + r^2 \tau_1(t) \qquad (2)$$

Here the crossover time of the $\tau_1(t)$ term is determined by the rotational diffusion coefficient, $\tau_* = 1/(2D_\theta) = 11$ s. Eq. 2 indicates that the short-time diffusion coefficient, $\overline{D}^{ST} = \overline{D}^{CoH} + r^2 D_\theta/2$, is dependent on the position of the TP, while the long time diffusion coefficient, $\overline{D}^{LT} = \overline{D}^{CoH}$, is independent of the TP. This expression fits very well the experimental data [Fig. 2(a)]. This crossover has been predicted by previous theory [25, 46] and is observed for the first time in experiments.

Our model also provides a clear physical picture of the nonzero MDs and the crossover between short- and long-time diffusion. As the rotational Brownian motion produces random displacements of the CoB on an arc with the CoH as its center, the projected displacements are symmetric to the CoB along $X_2$ direction, while biased towards the arc center (i.e, the CoH) along $X_1$ direction. When this Brownian orbital motion of the CoB covers a circle for $t \gg \tau_\theta$, the MDs saturate and the Brownian motion crosses over to the long-time diffusion.

To measure the other elements of the diffusion tensor, the translational displacements need to be transformed into a body frame co-moving with the particle. One convenient body frame has its origin fixed at the CoB and $X_1$ axis coincident with the symmetry axis [Fig. 1(b)]. The displacements between consecutive body frames were obtained from those in the lab frame through the rotational transformation $\Delta X_i(\tau, t_n) = R_{ij}(\theta_n) \Delta x_j(\tau, t_n)$, where $i, j = 1$ or 2, and $R_{ij}(\theta_n)$ is the rotation transformation matrix. The body frame trajectories are constructed by accumulating the displacements, $X_i(t_n) = \sum_{k=0}^{n} \Delta X_i(t_k)$.

One has two different choices of $\theta_n$: $\theta_n = \theta(t_n)$ representing the orientation at the beginning of each time interval, or $[\theta(t_n) + \theta(t_{n+1})]/2$ representing the average orientation during the time interval. One previous work shows that these two choses give indistinguishable results for particles of high symmetry like ellipsoids [5]. However, we find that the distinction between these two frames becomes important for low-symmetry particles such as the boomerangs. We term the first $[\theta_n = \theta(t_n)]$ as the discrete body frame (DBF) and the second $[\theta_n = [\theta(t_n) + \theta(t_{n+1})]/2]$ as the continuous body frame (CBF) [45].

In the CBF, the measured MDs along both $X_1$ and $X_2$ directions are zero [Fig. 3(a)], and the MSDs are linear with time, $\left\langle [\Delta X_i(t)]^2 \right\rangle = 2D_{ii}t$ with the diffusion coefficients $D_{11} = 0.049$ µm²/s and $D_{22} = 0.117$ µm²/s [Fig. 3(c)]. In contrast, very different behaviors are observed in the DBF. While the MD along $X_2$ is zero, the MD along $X_1$ is nonzero and grows linearly with time [Fig. 3(b)]. Accompanying this nonzero drift, the MSDs along $X_1$ exhibit nonlinearity with time [Fig. 3(d)]. Since the orientation of the DBF is reset at the beginning of each time step, the non-zero value of $\left\langle \Delta X_1 \right\rangle$ seen in the DBF is actually a manifestation of the non-zero MDs $\left\langle \Delta x_1 \right\rangle$ observed in the lab frame for $\theta_0 = 0$.

Since the displacements of the CoB along $X_2$ tend to induce rotation and vice versa, the coupled diffusion coefficient $D_{2\theta}$ is nonzero. While the translational motion of the CoB along $X_1$ is decoupled with rotation, $D_{1\theta}$ is thus zero. Our experimental results show that the DBF and CBF give rise to similar results for the translation-rotation correlation functions [Fig. 3 (e, f)]. Linear fitting of these data with $\langle \Delta X_i(t) \Delta \theta(t) \rangle = 2D_{i\theta}t$ gives a negligible value for $D_{1\theta}$ ($\leq$ -0.002 μm·rad/s), and $D_{2\theta}$ = 0.051 μm·rad/s [Fig. 3 (e, f)].

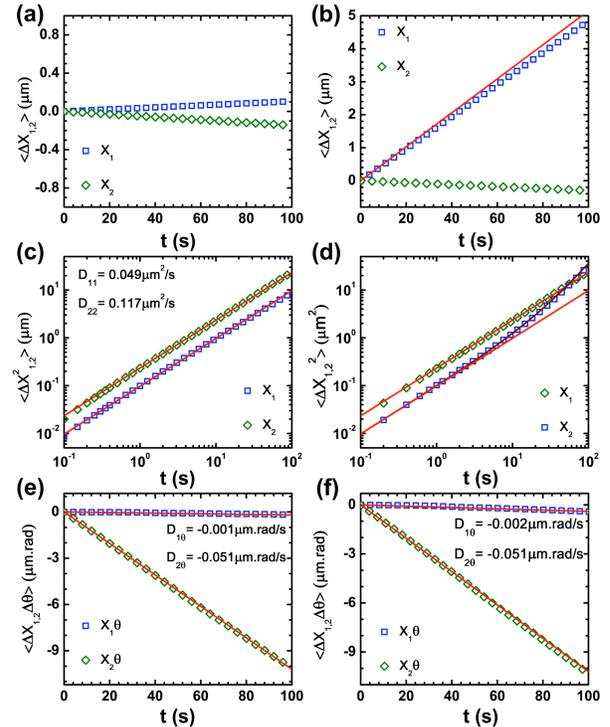

**Fig. 3.** (a-b) MDs vs. $t$ in the CBF (a) and DBF (b). The red line in (b) is the theory curve of Eq. 5a. (c-d) MSDs vs. $t$ in the CBF (c) and DBF (d). In (c), the red lines are the best linear fittings. In (d), the dark brown curve is Eq. 5b with $D_\theta$ and $D_{2\theta}$ obtained from the data in Fig. 2 and the red straight lines have same slopes as those in (c). (e-f) Translation-rotation correlations vs. $t$ in the CBF (e) and DBF (f). The red lines are the best linear fitting.

The differences between the CBF and DBF can be ascribed to the differences between the displacements of the vector **r** in these two body frames. The first and second moments of displacements for the TP can be expressed in the CBF as [45]:

$$\langle \Delta X_1(t) \rangle = \langle \Delta X_2(t) \rangle = 0 \quad (4a)$$
$$\langle \Delta X_1^2(t) \rangle = 2D_{11}t = 2D_{11}^{CoH}t \quad (4b)$$
$$\langle \Delta X_2^2(t) \rangle = 2D_{22}t = 2(D_{22}^{CoH} + r^2 D_\theta)t \quad (4c)$$

and in the DBF as:

$$\langle \Delta X_1(t) \rangle = rD_\theta t; \quad \langle \Delta X_2(t) \rangle = 0 \quad (5a)$$
$$\langle \Delta X_1^2(t) \rangle = 2D_{11}^{CoH}t + (rD_\theta)^2 t^2 \quad (5b)$$
$$\langle \Delta X_2^2(t) \rangle = 2D_{22}t = 2(D_{22}^{CoH} + r^2 D_\theta)t \quad (5c)$$

The CBF and DBF give the same forms for the translation-rotation correlation functions: $\langle \Delta X_1 \Delta \theta \rangle = 0$, and $\langle \Delta X_2 \Delta \theta \rangle = 2D_{2\theta}t = 2rD_\theta t$ [45].

Based on Eq. 1a and Eq. 5a, the slopes of the MDs vs. time are the same at short times in the DBF and in the lab frame (for $\theta_0 = 0$), which agrees with the experiments [Fig. 2d, 3b]. Employing the DBF provides a physical picture consistent with the lab frame observations, while using the CBF averages out the drift term in MDs and the nonlinear components in MSDs and provides a convenient way to calculate diffusion coefficients.

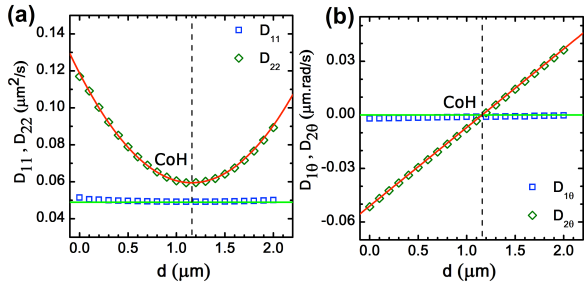

**Fig. 4.** (a) $D_{11}$ and $D_{22}$ vs. *d*. Red and green lines are theory curves based on Eq. 5(b, c). (b) $D_{1\theta}$ and $D_{2\theta}$ vs. *d*. Red and green lines are the theory curves of $D_{2\theta} = rD_\theta$ and $D_{1\theta}=0$ respectively. The dashed lines indicate the CoH at $d_0$ = 1.16 μm.

To verify the existence of the CoH, we re-calculated the trajectories and the diffusion coefficients for TPs on the symmetry line which have different distance *d* from the CoB, here $d=d_0-r$. $D_{22}$ and $D_{2\theta}$ can be derived from the above equations as a function of *r*: $D_{22} = D_{22}^{CoH} + r^2 D_\theta$, and $D_{2\theta} = rD_\theta$. We see that the theoretical curves fit the experimental results well [Fig. 4(a, b)]: $D_{11}$ remains unchanged, $D_{22}$ reaches a minimum at $d$ = 1.16 μm and $D_{2\theta}$ increases with *d* and crosses zero approximately at 1.16μm. These indicate that the CoH is at a distance $d_0$=1.16 μm from the CoB, which agrees with $d_0 = D_{2\theta}/D_\theta$.

With the CoH as TP, the MSDs in the lab and body frames all grow linearly with time and the translation-rotation correlation functions are zero [Fig. S1(a-c)]. As expected, the differences between DBF and CBF disappear. The averaged diffusion coefficient for the CoH in the lab frame, $\overline{D}^{CoH}$ = 0.054 μm$^2$/s, agrees with the long-time averaged diffusion coefficient of the CoB.

It is not trivial to validate the existence of the CoH for the boomerangs in quasi-2D confinements since previous theory shows that the Brownian motion of bent-rods in three dimensions exhibit screw-like properties [25]. Experiments with 1.7 μm and 1.9 μm thick cells for the same or different boomerang particles show qualitatively similar results with different values of diffusion coefficients. Cell confinements are known to significantly affect the diffusion coefficients [47-49]. It is worth to further explore in the future how the Brownian motion of the boomerangs crosses over from the non-skewed 2D behaviors to skewed 3D behaviors when the cell thickness is raised.

In conclusion, we have shown that the Brownian motion of the boomerang particles exhibits non-zero MDs for fixed initial angles, TP-dependent short-time and TP-independent long-time diffusion coefficients as results of non-overlap between the TP and its CoH. These results observed in boomerangs should occur in any non-skewed particles as long as the TP is not coincident with the CoH, and thus have important implications for studying the diffusion and transport of anisotropic particles. The subtle difference between the CBF and DBF has significant influence on proper calculations of diffusion coefficients. Prior theoretical studies have provided analytical results regarding the orientation distributions of bent-rods under external fields [39-42]; it will be interesting to study how the boomerangs transport under gravitational or electrophoretic forces.

The authors would like to acknowledge Hartmut Löwen, Tom Lubensky, and Kun Zhao for valuable discussions and Oleg Lavrentovich for both valuable discussions and using the equipment in his lab. Partial financial support from NSF Grant DMR-1106014, ECCS-0824175 and Farris Family Award are acknowledged.

*Correspondence to: qwei@kent.edu, jselinge@kent.edu.

**Supplementary Material for "Brownian Motion of Boomerang Colloidal Particles"**

Ayan Chakrabarty[1], Andrew Konya[1], Feng Wang[1], Jonathan V. Selinger[1]*, Kai Sun[2], Qi-Huo Wei[1]*
[1]Liquid Crystal Institute, Kent State University, Kent, OH 44242
[2]Department of Materials Science and Engineering, University of Michigan, Ann Arbor, MI 48109, USA

**Theoretical Derivations**

**1. Displacements of the tracking point on the symmetric line of the boomerang**

As pointed out by Brenner and others [22, 25, 27], the center of hydrodynamic stress (CoH) is located on particle's symmetric lines. Assuming that the CoH for the boomerang is located at $(x_1^{CoH}, x_2^{CoH})$ on the bisector of the apex angle, then the position of a tracking point (TP) on the symmetry line is simply the sum of the CoH position vector and the vector $\mathbf{r} = -r\cos\theta \hat{x}_1 - r\sin\theta \hat{x}_2$ from the CoH to the TP:

$$\begin{pmatrix} x_1(t) \\ x_2(t) \end{pmatrix} = \begin{pmatrix} x_1^{CoH}(t) \\ x_2^{CoH}(t) \end{pmatrix} - r \begin{pmatrix} \cos\theta(t) \\ \sin\theta(t) \end{pmatrix} \tag{S1}$$

where $\theta(t)$ is the orientation of the particle. Therefore the motion of the TP can be described by:

$$\begin{pmatrix} \dot{x}_1(t) \\ \dot{x}_2(t) \end{pmatrix} = \begin{pmatrix} \dot{x}_1^{CoH}(t) \\ \dot{x}_2^{CoH}(t) \end{pmatrix} + r \begin{pmatrix} \sin\theta(t) \\ -\cos\theta(t) \end{pmatrix} \dot{\theta}(t) \tag{S2}$$

Since the displacements of $\mathbf{x}^{CoH}$ and $\mathbf{r}$ are not correlated, i.e., $\langle [\Delta x_i^{CoH}(t)][\Delta r_i(t)] \rangle = 0$, the mean displacements (MDs) and mean square displacements (MSDs) of the TP can be written as:

$$\langle \Delta x_i(t) \rangle = \langle \Delta x_i^{CoH}(t) \rangle + \langle \Delta r_i(t) \rangle \tag{S3}$$

$$\langle [\Delta x_i(t)]^2 \rangle = \langle [\Delta x_i^{CoH}(t)]^2 \rangle + \langle [\Delta r_i(t)]^2 \rangle \tag{S4}$$

where $i = 1, 2$.

**1.1 Displacements of the CoH**

According to the definition of the CoH, the diffusion tensor and the resistance tensor for the CoH are diagonalized. Therefore, under over-damped conditions, the inertial term in the Langevin equation can be neglected, and the Langevin Equations for the CoH under no external force in the body frame can be written in 2D as:

$$\begin{pmatrix} \zeta_{11}^{CoH} & 0 & 0 \\ 0 & \zeta_{22}^{CoH} & 0 \\ 0 & 0 & \zeta_{\theta\theta} \end{pmatrix} \begin{pmatrix} \dot{X}_1^{CoH}(t) \\ \dot{X}_2^{CoH}(t) \\ \dot{\theta}(t) \end{pmatrix} = \begin{pmatrix} \xi_1(t) \\ \xi_2(t) \\ \xi_\theta(t) \end{pmatrix} \tag{S5}$$

where $\zeta_{ij}^{CoH}$ is the hydrodynamic resistance tensor. The Gaussian random noise $\xi_i(t)$ is related to the resistance tensor through the fluctuation-dissipation theorem:

$$\langle \xi_i(t) \rangle = 0$$
$$\langle \xi_i(t)\xi_j(t') \rangle = 2k_B T \zeta_{ij}^{CoH} \delta(t-t')$$

Here $i, j = 1, 2, \theta$. The diffusion tensor follows the generalized Einstein-Smoluchowski relationship: $D_{ij}^{CoH} = k_B T (\zeta_{ij}^{CoH})^{-1}$. Eq. S5 can be rewritten as $\dot{X}_i^{CoH} = \frac{1}{k_B T} D_{ij}^{CoH} \xi_j(t)$. For simplification, we scale the random noise as: $\xi_j(t) = k_B T \sqrt{2/D_{ij}^{CoH}} \eta_j(t)$, where $\eta_i(t)$ is a random variable with a normal Gaussian distribution, i.e. $\langle \eta_i(t) \rangle = 0; \langle \eta_i(t)\eta_j(t') \rangle = \delta_{ij}\delta(t-t')$. The Langevin Equation can then be written separately for the translational and rotational components as:

$$\begin{pmatrix} \dot{X}_1^{CoH}(t) \\ \dot{X}_2^{CoH}(t) \end{pmatrix} = \begin{pmatrix} \sqrt{2D_{11}^{CoH}} & 0 \\ 0 & \sqrt{2D_{22}^{CoH}} \end{pmatrix} \begin{pmatrix} \eta_1(t) \\ \eta_2(t) \end{pmatrix} \tag{S6a}$$

$$\dot{\theta}(t) = \sqrt{2D_\theta}\eta_\theta(t) \tag{S6b}$$

The equations of motion in the lab frame are obtained through a rotation transformation of the body frame Eq. S6a:

$$\begin{pmatrix} \dot{x}_1^{CoH}(t) \\ \dot{x}_2^{CoH}(t) \end{pmatrix} = \begin{pmatrix} \cos\theta(t) & -\sin\theta(t) \\ \sin\theta(t) & \cos\theta(t) \end{pmatrix} \begin{pmatrix} \sqrt{2D_{11}^{CoH}} & 0 \\ 0 & \sqrt{2D_{22}^{CoH}} \end{pmatrix} \begin{pmatrix} \eta_1(t) \\ \eta_2(t) \end{pmatrix} \tag{S7}$$

These Langevin equations S6-S7 are the same as those for an ellipsoidal particle whose solutions have been given in Ref. [5]. It can be easily seen that the MDs of the CoH for fixed initial angle $\theta_0$ are zero:

$$\langle \Delta x_1^{CoH}(t) \rangle_{\theta_0} = \langle \Delta x_2^{CoH}(t) \rangle_{\theta_0} = 0 \tag{S8}$$

Using the results in Ref. [5], the MSDs of the CoH with fixed initial orientation $\theta_0$ take the form:

$$\langle [\Delta x_1^{CoH}(t)]^2 \rangle_{\theta_0} = \cos^2\theta_0 \left[ (D_{11}^{CoH} + D_{22}^{CoH})t - \frac{\Delta D^{CoH}}{4D_\theta}(1-e^{-4D_\theta t}) \right]$$
$$+ \sin^2\theta_0 \left[ (D_{11}^{CoH} + D_{22}^{CoH})t + \frac{\Delta D}{4D_\theta}(1-e^{-4D_\theta t}) \right] \tag{S9a}$$

$$\langle [\Delta x_2^{CoH}(t)]^2 \rangle_{\theta_0} = \cos^2\theta_0 \left[ (D_{11}^{CoH} + D_{22}^{CoH})t + \frac{\Delta D^{CoH}}{4D_\theta}(1-e^{-4D_\theta t}) \right]$$
$$+ \sin^2\theta_0 \left[ (D_{11}^{CoH} + D_{22}^{CoH})t - \frac{\Delta D}{4D_\theta}(1-e^{-4D_\theta t}) \right] \tag{S9b}$$

where $\Delta D = D_{22}^{CoH} - D_{11}^{CoH}$

### 1.2 Displacements of vector *r*

The displacements of *r* along the $x_1$ and $x_2$ directions can be expressed as:

$$\Delta r_1(t) = r\cos\theta_0 - r\cos[\theta_0 + \Delta\theta(t)] \tag{S10a}$$
$$\Delta r_2(t) = r\sin\theta_0 - r\sin[\theta_0 + \Delta\theta(t)] \tag{S10b}$$

Given that $\Delta\theta(t)$ has a Gaussian distribution with a zero mean, it can be shown that for integer $n$, $\langle\sin[n\Delta\theta(t)]\rangle = 0$, and $\langle\cos[n\Delta\theta(t)]\rangle = \exp(-n^2 D_\theta t)$. The MDs of the vector $r$ then become:

$$\langle\Delta r_1(t)\rangle_{\theta_0} = r\cos\theta_0\left(1 - e^{-D_\theta t}\right) \tag{S11a}$$
$$\langle\Delta r_2(t)\rangle_{\theta_0} = r\sin\theta_0\left(1 - e^{-D_\theta t}\right) \tag{S11b}$$

The MSDs of $r$ re derived as the following:

$$\langle[\Delta r_1(t)]^2\rangle_{\theta_0} = r^2\cos^2\theta_0 - 2r^2\cos\theta_0\langle\cos[\theta_0 + \Delta\theta(t)]\rangle + \frac{r^2}{2}\langle 1 + \cos 2[\theta_0 + \Delta\theta(t)]\rangle$$

Or
$$\langle[\Delta r_1(t)]^2\rangle_{\theta_0} = \frac{1}{2}r^2\cos^2\theta_0\left(3 - 4e^{-D_\theta t} + e^{-4D_\theta t}\right) + \frac{1}{2}r^2\sin^2\theta_0\left(1 - e^{-4D_\theta t}\right) \tag{S12a}$$

Similarly,
$$\langle[\Delta r_2(t)]^2\rangle_{\theta_0} = \frac{1}{2}r^2\sin^2\theta_0\left(3 - 4e^{-D_\theta t} + e^{-4D_\theta t}\right) + \frac{1}{2}r^2\cos^2\theta_0\left(1 - e^{-4D_\theta t}\right) \tag{S12b}$$

### 1.3 MD and MSD of the Tracking Point

Substituting Eq. S8 and Eq. S11 (a-b) into Eq. S3 yields the MDs of the TP with fixed initial angle $\theta_0$:

$$\langle\Delta x_1(t)\rangle_{\theta_0} = r\cos\theta_0\left(1 - e^{-D_\theta t}\right) \tag{S13a}$$
$$\langle\Delta x_2(t)\rangle_{\theta_0} = r\sin\theta_0\left(1 - e^{-D_\theta t}\right) \tag{S13b}$$

Substituting Eq. 9 (a-b) and Eq. S12 (a-b) into Eq. S4 yields the MSDs of the TP for fixed initial orientation $\theta_0$:

$$\langle[\Delta x_1(t)]^2\rangle_{\theta_0} = \cos^2\theta_0\left[2\overline{D}^{CoH}t - \frac{\Delta D}{4D_\theta}\left(1 - e^{-4D_\theta t}\right) + \frac{r^2}{2}\left(3 - 4e^{-D_\theta t} + e^{-4D_\theta t}\right)\right] + $$
$$\sin^2\theta_0\left[2\overline{D}^{CoH}t + \frac{\Delta D}{4D_\theta}\left(1 - e^{-4D_\theta t}\right) + \frac{r^2}{2}\left(1 - e^{-4D_\theta t}\right)\right] \tag{S14a}$$

$$\langle[\Delta x_2(t)]^2\rangle_{\theta_0} = \sin^2\theta_0\left[2\overline{D}^{CoH}t - \frac{\Delta D}{4D_\theta}\left(1 - e^{-4D_\theta t}\right) + \frac{r^2}{2}\left(3 - 4e^{-D_\theta t} + e^{-4D_\theta t}\right)\right] + $$
$$\cos^2\theta_0\left[2\overline{D}^{CoH}t + \frac{\Delta D}{4D_\theta}\left(1 - e^{-4D_\theta t}\right) + \frac{r^2}{2}\left(1 - e^{-4D_\theta t}\right)\right] \tag{S14b}$$

where $\overline{D}^{CoH} = (D_{11}^{CoH} + D_{22}^{CoH})/2$.

Specifically, when the CoB is used as the TP with $\theta_0 = 0$, and $r$ represents the distance of the CoH from the CoB, the MDs are simplified as:

$$\langle\Delta x_1(t)\rangle_{\theta_0=0} = r\left(1 - e^{-D_\theta t}\right) \tag{S15a}$$
$$\langle\Delta x_2(t)\rangle_{\theta_0=0} = 0 \tag{S15b}$$

$$\left\langle [\Delta x_1(t)]^2 \right\rangle_{\theta_0=0} = 2\overline{D}^{CoH} t - \left(\frac{\Delta D}{4D_\theta} + \frac{r^2}{2}\right)\left(1 - e^{-4D_\theta t}\right) + 2r^2\left(1 - e^{-D_\theta t}\right) \quad \text{(S16a)}$$

$$\left\langle [\Delta x_2(t)]^2 \right\rangle_{\theta_0=0} = 2\overline{D}^{CoH} t + \left(\frac{\Delta D}{4D_\theta} + \frac{r^2}{2}\right)\left(1 - e^{-4D_\theta t}\right) \quad \text{(S16b)}$$

Averaging Eq. S15 (a-b) and S16 (a-b) over different initial orientation $\theta_0$ yields the angle averaged MDs and MSDs in the lab frame:

$$\langle \Delta x_1(t) \rangle = \langle \Delta x_2(t) \rangle = 0 \quad \text{(S17)}$$

$$\left\langle [\Delta x_1(t)]^2 \right\rangle = \left\langle [\Delta x_2(t)]^2 \right\rangle = 2\overline{D}^{CoH} t + r^2\left(1 - e^{-D_\theta t}\right) \quad \text{(S18)}$$

Eq. S18 indicates that the averaged diffusion coefficient of the CoB in the lab frame is $\overline{D}^{ST} = \overline{D}^{CoH} + \frac{1}{2}r^2 D_\theta$ for short times and $\overline{D}^{LT} = \overline{D}^{CoH}$ for long time.

## 2. The Two Body Frames

Since the displacements obtained from the trajectories are in the lab frame, a rotation transformation needs to be performed to determine the elements of the diffusion tensor. We start with velocities in the body frame which are related to those in the lab frame through:

$$\dot{X}_1(t) = \cos\theta'(t)\dot{x}_1(t) + \sin\theta'(t)\dot{x}_2(t)$$
$$\dot{X}_2(t) = -\sin\theta'(t)\dot{x}_1(t) + \cos\theta'(t)\dot{x}_2(t) \quad \text{(S19)}$$

where $\theta'(t)$ is the angle used to transform the lab frame velocity (and displacements) into the body frame. To note, we use $\theta'(t)$ to distinguish it from $\theta(t)$. Using Eq. S2 yields:

$$\dot{X}_1(t) = \dot{X}_1^{CoH}(t) + \sin[\theta(t) - \theta'(t)]r\alpha_\theta\eta_\theta(t) = \dot{X}_1^{CoH}(t) + \dot{R}_1(t)$$
$$\dot{X}_2(t) = \dot{X}_2^{CoH}(t) - \cos[\theta(t) - \theta'(t)]r\alpha_\theta\eta_\theta(t) = \dot{X}_2^{CoH}(t) + \dot{R}_2(t) \quad \text{(S20)}$$

where $\dot{R}_1(t)$ and $\dot{R}_2(t)$ are the velocities of the vector **r** transformed into the body frame, and $\dot\theta(t) = \sqrt{2D_\theta}\eta_\theta(t) = \alpha_\theta\eta_\theta(t)$ from Eq. S6b. The MDs, MSDs and cross coupling of the TP in the body frame are simply:

$$\langle \Delta X_i(t) \rangle = \langle \Delta X_i^{CoH}(t) \rangle + \langle \Delta R_i(t) \rangle \quad \text{(S21)}$$

$$\left\langle [\Delta X_i(t)]^2 \right\rangle = \left\langle [\Delta X_i^{CoH}(t)]^2 \right\rangle + \left\langle [\Delta R_i(t)]^2 \right\rangle \quad \text{(S22)}$$

$$\langle \Delta X_i(t)\Delta\theta(t) \rangle = \langle \Delta X_i^{CoH}(t)\Delta\theta(t) \rangle + \langle \Delta R_i(t)\Delta\theta(t) \rangle \quad \text{(S23)}$$

In experiments, for the time interval between $t_n$ and $t_{n+1}$, one can choose either $\theta'(t) = [\theta(t_n) + \theta(t_{n+1})]/2$, which define a body frame noted here as the continuous body frame (CBF), or $\theta'(t) = \theta(t_n)$, which define another body frame noted here as the discrete body frame (DBF).

As shown in Ref. [5], these two body frames yield the same results for ellipsoidal particles. Since Eqs. S6 (a-b) are the same to those for ellipsoids, we have $\langle \Delta X_i^{CoH}(t) \rangle = 0$ and $\left\langle \left[ \Delta X_i^{CoH}(t) \right]^2 \right\rangle = 2 D_{ii}^{CoH} t$ where $i = 1, 2$. There is also no coupling with the rotational diffusion at CoH and hence $\left\langle \Delta X_i^{CoH}(t) \Delta \theta(t) \right\rangle = 0$.

In the following we show theoretically that for boomerangs these two body frames yield different results due to different displacements of the vector $\boldsymbol{r}$. This result implies that as long as the TP is not coincident with the CoH, these two body frames are different regardless of the particle shape.

## 2.1. Motion of $r$ in the Body Frame

From Eq. S20, we rewrite the velocities of $\boldsymbol{r}$ in the body frame as:
$$\dot{R}_1(t) = \sin[\theta(t) - \theta'(t)] r \alpha_\theta \eta_\theta(t) \tag{S24a}$$
$$\dot{R}_2(t) = -\cos[\theta(t) - \theta'(t)] r \alpha_\theta \eta_\theta(t) \tag{S24b}$$

For a time step between $t_0$ and $t_0+\tau$, the displacements of $R_1$ and $R_2$ are respectively $\Delta R_1(\tau) = \int_{t_0}^{t_0+\tau} dt' \sin[\theta(t') - \theta'(t')] r \alpha_\theta \eta_\theta(t')$, $\Delta R_2(\tau) = -\int_{t_0}^{t_0+\tau} dt' \cos[\theta(t') - \theta'(t')] r \alpha_\theta \eta_\theta(t')$.

Due to the discretized nature of the experimental trajectories, $\theta'(t')$ used in either the DBF or CBF is not the same as the instantaneous angle $\theta(t')$. For small $\tau$, $\theta(t') - \theta'(t')$ is small, and the MDs can be expressed as:

$$\langle \Delta R_1(\tau) \rangle = \left\langle \int_{t_0}^{t_0+\tau} dt' [\theta(t') - \theta'(t')] r \alpha_\theta \eta_\theta(t') \right\rangle \tag{S25}$$

$$\langle \Delta R_2(\tau) \rangle = -\left\langle \int_{t_0}^{t_0+\tau} dt' \, r \alpha_\theta \eta_\theta(t') \right\rangle = 0 \tag{S26}$$

the MSDs of $R_1$ and $R_2$ can be expressed as:

$$\left\langle [\Delta R_1(\tau)]^2 \right\rangle = r^2 \alpha_\theta^2 \left\langle \int_{t_0}^{t_0+\tau} dt' [\theta(t') - \theta'(t')] \eta_\theta(t') \int_{t_0}^{t_0+\tau} dt'' [\theta(t'') - \theta'(t'')] \eta_\theta(t'') \right\rangle, \tag{S27}$$

$$\left\langle [\Delta R_2(\tau)]^2 \right\rangle = \left\langle \int_{t_0}^{t_0+\tau} dt' \, r \alpha_\theta \eta_\theta(t') \int_{t_0}^{t_0+\tau} dt'' \, r \alpha_\theta \eta_\theta(t'') \right\rangle, \tag{S28}$$

and the coupling with rotation for $R_1$ and $R_2$ can be expressed as:

$$\langle \Delta R_1(\tau) \Delta \theta(\tau) \rangle = r^2 \alpha_\theta^2 \left\langle \int_{t_0}^{t_0+\tau} dt' [\theta(t') - \theta'(t')] \eta_\theta(t') \int_{t_0}^{t_0+\tau} dt'' \eta_\theta(t'') \right\rangle, \tag{S29}$$

$$\langle \Delta R_2(\tau) \Delta \theta(\tau) \rangle = \left\langle \int_{t_0}^{t_0+\tau} dt' \, r \alpha_\theta \eta_\theta(t') \int_{t_0}^{t_0+\tau} dt'' \, \alpha_\theta \eta_\theta(t'') \right\rangle \tag{S30}$$

For a longer time $t = n\tau$ with $n$ being an integer, the displacements of $\boldsymbol{r}$ in the body frame are obtained by accumulating the displacements in individual time steps, i.e., $\Delta R_{1,2}(t,t_0) = \sum_{i=0}^{n-1} \Delta R_{1,2}(\tau, t_0 + i\tau)$. The MDs, MSDs and the cross coupling displacements are then given by

$$\langle \Delta R_{1,2}(t) \rangle = \left\langle \sum_{i=0}^{n-1} \Delta R_{1,2}(\tau, t_0 + i\tau) \right\rangle = n \langle \Delta R_{1,2}(\tau) \rangle \tag{S31a}$$

$$\begin{aligned}\langle [\Delta R_{1,2}(t)]^2 \rangle &= \left\langle \left[\sum_{i=0}^{n-1} \Delta R_{1,2}(\tau, t_0 + i\tau)\right]\left[\sum_{j=0}^{n-1} \Delta R_{1,2}(\tau, t_0 + j\tau)\right] \right\rangle \\ &= \left\langle \sum_{i=0}^{n-1}[\Delta R_{1,2}(\tau, t_0 + i\tau)]^2 \right\rangle + \left\langle \sum_{i \neq j}[\Delta R_{1,2}(\tau, t_0 + i\tau)][\Delta R_{1,2}(\tau, t_0 + j\tau)] \right\rangle \\ &= n\langle [\Delta R_{1,2}(\tau)]^2 \rangle + (n^2 - n)\langle [\Delta R_{1,2}(\tau)][\Delta R_{1,2}(\tau)] \rangle \end{aligned} \tag{S31b}$$

$$\langle \Delta R_{1,2}(t)\Delta\theta(t) \rangle = \left\langle \left[\sum_{i=0}^{n-1} \Delta R_{1,2}(\tau, t_0 + i\tau)\right]\left[\sum_{j=0}^{n-1} \Delta\theta(\tau, t_0 + j\tau)\right] \right\rangle = n\langle \Delta R_{1,2}(\tau)\Delta\theta(\tau) \rangle \tag{S31c}$$

### 2.1.1. Motion of *r* in the Continuous Body Frame (CBF)

For the time interval between $t_0$ and $t_0+\tau$ in the CBF, $\theta'(t') = [\theta(t_0) + \theta(t_0 + \tau)]/2$. Using the expressions $\theta(t_0) = \theta(t') - \int_{t_0}^{t'} dt'' \dot{\theta}(t'')$ and $\theta(t_0 + \tau) = \theta(t') + \int_{t'}^{t_0+\tau} dt'' \dot{\theta}(t'')$, we have

$$\theta'(t') = \theta(t') + \frac{\alpha_\theta}{2}\left[\int_{t'}^{t_0+\tau} dt'' \eta_\theta(t'') - \int_{t_0}^{t'} dt'' \eta_\theta(t'')\right] \tag{S32}$$

Substituting S32 into Eq. S25, S27 and Eq. S29 lead to:
$$\langle \Delta R_1(\tau) \rangle = 0 \tag{S33}$$
$$\langle [\Delta R_1(\tau)]^2 \rangle = 0 \tag{S34}$$
$$\langle \Delta R_1(\tau)\Delta\theta(t) \rangle = 0 \tag{S35}$$

Based on Eq. S28 and S30, it can be shown that
$$\langle [\Delta R_2(\tau)]^2 \rangle = 2r^2 D_\theta \tau \tag{S36}$$
$$\langle \Delta R_2(\tau)\Delta\theta(t) \rangle = 2rD_\theta \tau \tag{S37}$$

Combining Eq. S21, S22, S23, S26, S31a, S31b, S31c, S33, S34, S35, S36 and S37 leads to MDs, MSDS and cross coupling displacements of the TP in the CBF:
$$\langle \Delta X_1(t) \rangle = 0 \tag{S38}$$
$$\langle \Delta X_2(t) \rangle = 0 \tag{S39}$$
$$\langle [X_1(t)]^2 \rangle = 2D_{11}^{CoH} t \tag{S40}$$
$$\langle [X_2(t)]^2 \rangle = 2(D_{22}^{CoH} + r^2 D_\theta)t \tag{S41}$$
$$\langle X_1(t)\Delta\theta(t) \rangle = 0 \tag{S42}$$
$$\langle X_2(t)\Delta\theta(t) \rangle = 2rD_\theta t \tag{S43}$$

### 2.1.2. Motion of *r* in the Discrete Body Frame (DBF)

In the DBF, $\theta'(t')$ for the time interval between $t_0$ and $t_0+\tau$ in S24 and S26 can be expressed as:

$$\theta'(t') = \theta(t_0) = \theta(t') - \int_{t_0}^{t'} dt'' \dot{\theta}(t'') \tag{S44}$$

Substituting Eq. S44 in Eq. S25 gives:

$$\langle \Delta R_1(\tau) \rangle = r\alpha_\theta^2 \int_{t_0}^{t_0+\tau} dt' \int_{t_0}^{t'} dt'' \langle \eta_\theta(t')\eta_\theta(t'') \rangle$$

$$= 2rD_\theta \int_{t_0}^{t_0+\tau} dt' \int_{t_0}^{t'} dt'' \delta(t'-t'') \tag{S45}$$

$$= rD_\theta \tau$$

Using Eq. S44, Eq. S27 can be evaluated as:

$$\langle \Delta R_1^2(\tau) \rangle = r^2\alpha_\theta^2 \left\langle \int_{t_0}^{t_0+\tau} dt' \eta_\theta(t') \int_{t_0}^{t'} dt'' \dot{\theta}(t'') \int_{t_0}^{t_0+\tau} dt''' \eta_\theta(t''') \int_{t_0}^{t'} dt'''' \dot{\theta}(t'''') \right\rangle \tag{S46}$$

$$= 3r^2 D_\theta^2 \tau^2$$

Using Eq. S44, Eq. S29 can be evaluated as:

$$\langle \Delta R_1(\tau) \Delta\theta(t) \rangle = 0 \tag{S47}$$

Based on Eq. S28 and S30 the MSD the cross coupling of $R_2$ in the DBF is the same as in the CBF:

$$\langle \Delta R_2^2(\tau) \rangle = 2r^2 D_\theta \tau \tag{S48}$$

$$\langle \Delta R_2(\tau) \Delta\theta(t) \rangle = 0 \tag{S49}$$

Combining Eq. S21, S22, S23, S26, S31a, S31b, S31c, S45, S46, S47, S48 and S49 leads to MDs, MSDS and cross coupling and of the TP in the DBF:

$$\langle \Delta X_1(t) \rangle = rD_\theta t \tag{S41}$$

$$\langle \Delta X_2(t) \rangle = 0 \tag{S42}$$

$$\langle [X_1(t)]^2 \rangle = 2D_{11}^{CoH} t + r^2 D_\theta^2 t^2 \tag{S43}$$

$$\langle [X_2(t)]^2 \rangle = 2(D_{22}^{CoH} + r^2 D_\theta) t \tag{S44}$$

$$\langle X_1(t) \Delta\theta(t) \rangle = 0 \tag{S42}$$

$$\langle X_2(t) \Delta\theta(t) \rangle = 2rD_\theta t \tag{S43}$$

**Supporting Figure**:

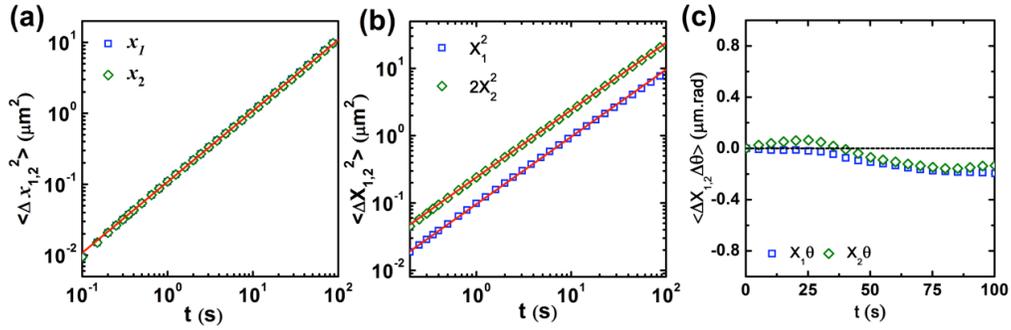

**Fig. S1**. (a) Angle averaged MSDs for the CoH measured in the lab frame. The red line is the best linear fit with $\bar{D}^{CoH}=0.054\,\mu m^2/s$. (b) MSDs of the CoH in the DBF. Red lines represent the best linear fitting with $D_{11}^{CoH} = 0.049\,\mu m^2/s$ and $D_{22}^{CoH} = 0.060\,\mu m^2/s$. For clarity, MSDs for $X_2$ are shifted by a factor of 2. (c) Translational and rotational displacement correlation functions calculated at CoH.